\documentclass[oupdraft]{bio}
\usepackage{url}

\history{Plan to submit November, 2017}

\begin{document}

\title{Evaluation of Treatment Effect Modification by Biomarkers Measured Pre- and Post-randomization in the Presence of Non-monotone Missingness}

\author{YINGYING ZHUANG$^\ast$, YING HUANG, PETER B. GILBERT\\[4pt]
\textit{Department of Biostatistics,
University of Washington, Seattle, WA, USA}
\\[2pt]
{yyzhuang@uw.edu}}

\markboth%
{Zhuang and others}
{Treatment Effect Modification with Baseline Sub-sampling}

\maketitle

\footnotetext{To whom correspondence should be addressed.}

\begin{abstract}
{In vaccine studies, investigators are often interested in studying effect modifiers of clinical treatment efficacy by biomarker-based principal strata, which is useful for selecting biomarker study endpoints for evaluating treatments in new trials, exploring biological mechanisms of clinical treatment efficacy, and studying mediators of clinical treatment efficacy. However, in trials where participants may enter the study with prior exposure therefore with variable baseline biomarker values, clinical treatment efficacy may depend jointly on a biomarker measured at baseline and measured at a fixed time after vaccination. Therefore, it is of interest to conduct a bivariate effect modification analysis by biomarker-based principal strata and baseline biomarker values. Previous methods allow this assessment if participants who have the biomarker measured at the the fixed time point post randomization would also have the biomarker measured at baseline. However, additional complications in study design could happen in practice. For example, in the Dengue correlates study, baseline biomarker values were only available from a fraction of participants who have biomarkers measured post-randomization. How to conduct the bivariate effect modification analysis in these studies remains an open research question. In this article, we propose an estimated likelihood method to utilize the sub-sampled baseline biomarker in the effect modification analysis and illustrate our method with datasets from two dengue phase 3 vaccine efficacy trials.}
\end{abstract}

\section{Introduction}
\label{sec1}
A common problem of interest within a randomized clinical trial is the evaluation of an inexpensive intermediate study endpoint, typically a biomarker, as an effect modifier of the clinical treatment efficacy which can accelerate research to apply and develop effective treatments against the clinical outcome. Motivated by randomized placebo-controlled vaccine efficacy trials, Gilbert and Hudgens \citet{gilbert2008evaluating}, henceforth GH, proposed a clinically relevant causal estimand called the Causal Effect Predictiveness (CEP) surface, using the principal stratification framework developed by \citet{frangakis2002principal} to assess whether and how treatment efficacy varies by subgroups defined by intermediate response endpoint principal strata. Procedures were developed for contrasting clinical risks under two treatment arms conditional on the pair of potential biomarker values under two treatment arms. The CEP surface assesses a causal effect of vaccine because the comparison groups are selected based on principal stratification, which is not subject to post-randomization selection bias. GH expressed the concept that studying the whole CEP surface is important that a more useful biomarker will have wide variability in the CEP surface thus is a strong effect modifier. Unfortunately, if no further assumptions are made, the CEP surface can not identified by the observed data because of the missing potential outcomes. \citet{follmann2006augmented} proposed two augmented trial design to solve the problem: baseline immunogenicity predictors (BIP) and closeout placebo vaccination (CPV). The BIP design uses baseline predictor(s) to infer the unobserved potential biomarker values, while the CPV design vaccinates placebo recipients who stay uninfected at the end of the follow-up and measures their immune response values, which are used in place of their biomarker values under vaccine. Examples of good baseline predictors are baseline biomarker measurements in trials where participants may enter the study with prior exposure and biomarker measurements at baseline reflects natural immunity arising from pre-trial exposure to the disease-causing pathogen. Furthermore, some baseline predictors may modify the CEP surface and contrasting clinical risks under each treatment assignment may depend jointly on those baseline predictors and the biomarker values measured at a fixed time after randomization. Therefore, it is of interest to estimate CEP in those baseline predictors defined subgroups. For example, an common question that emerges from dengue Phase III trials is whether it is the new biomarker response generated by the vaccine over the baseline value or the absolute biomarker value achieved following vaccination that predicts clinical treatment efficacy. Comparing the CEP surface within the baseline seropositive subgroup (defined as baseline biomarker value equal or above detection limit) to the baseline seronegative subgroup (defined as baseline biomarker value below detection limit) could provide important insights to this question. Previous methods allow this assessment if baseline predictors are measured in everyone. \citet{huang2017evaluating} studied a three-phase sampling design in which immune response is further measured among a subset of participants for whom the baseline predictors are available. However, additional complications in study design could happen in practice. For example, in dengue Phase III trials, the biomarker values at baseline were only measured in a fraction of those with the biomarker measured at the post-randomization time point. Our goal in this manuscript is to propose methods for a bivariate treatment effect modification analysis by biomarker-based principal strata and baseline covariates in general settings without requirements of a nested sub-sampling relationship between the immune response biomarker and baseline predictors, in other words, in the presence of non-monotone missingness, applicable to both the BIP-only design and the BIP+CPV design.
 
The remainder of this article is organized as follows. In Section 2, we introduce the problem setting and propose an estimator for the CEP curve that accounts for effects of the baseline covariates together with the vaccine-induced biomarker in the risk model. In Section 3, we evaluate the finite-sample performance of the proposed estimator through extensive numerical studies. In Section 4, we present an analysis of two Phase 3 dengue vaccine efficacy trials using our proposed methods. Finally, Section 5 contains a discussion of our proposed methods, their applications in other settings, and possible areas of future research.

\section{Methods}
\label{sec2}

Consider a study in which $N$ subjects are independently and randomly selected from a given population of interest and are randomly assigned to either placebo or vaccine at baseline (time 0). Let $Z=1$ if subject is randomized to vaccine and $Z=0$ if subject is randomized to placebo. Let $Q$ be a vector of baseline covariates used for modeling disease risk and $Q$ can be partitioned into two part, $X$ and $B$. $X$ denotes baseline covariates recorded for everyone at baseline such as gender and country while $B$ denotes baseline covariates that are only available in a subset of the $N$ trial participants, such as baseline biomarker measurements. Trial participants are followed for the primary clinical endpoint for a predetermined period of time and let $Y$ be the indicator of clinical endpoint event during the study follow-up period. At some fixed time $\tau >0$ post randomization, an immune response endpoint, $S$, is measured. Because $S$ must be measured prior to disease to evaluate its treatment effect modification, availability of $S$ is conditional on remaining clinical endpoint free at time $\tau$ (denoted by $Y^{\tau}=0$). If clinical endpoint occurs in the time interval $\left [ 0, \tau \right ]$ ($Y^{\tau}=1$), then $S$ is undefined and we set $S=\ast$. In a CPV component is incorporated in the trial design, all or a fraction of placebo recipients who remain free of the clinical endpoint at the closeout of the trial are vaccinated and the immune response biomarker $S_c$ is measured at time $\tau$ after vaccination. In addition, we consider cases where $S$ is continuous and subject to "limit of detection" left censoring. The observable random variable $S\equiv max(S^*,c)$ where c is the limit of detection and $S^*$ has a continuous cdf with $Pr(S^*\leq c)>0$.  Similarly to $S$, $S_c\equiv max(S_c^*,c)$. If $B$ denotes the baseline biomarker measurements, then  observable random variable $B\equiv max(B^*,c)$ where $B^*$ has a continuous cdf with $Pr(B^*\leq c)>0$. Let $S(z)$,$S^*(z)$, $Y^{\tau}(z)$, $Y(z)$ be the potential outcomes if assigned treatment $z$, for $z=0$ or $1$. We consider a general sampling framework where baseline covariates $X$ and the clinical outcome data $Y$ and $Y^{\tau}$ are measured for everyone, sampling probability of $B$ depends on $X$ and $Z$, and sampling probability of $S(1)$ or $S_c$ depends on $Y$, $X$ and $Z$. 

The CEP surface is defined in terms of the clinical risks under each treatment assignment, 
$risk_z(s_1, s_0)\equiv P\left ( Y(z)=1|S(1)=s_1, S(0)=s_0, Y^\tau(1)= Y^\tau(0)= 0\right )$ for $z=0,1$. It conditions on the counterfactual pair $(S(0), S(1))$  which forms a principal stratification and can be considered as an unobserved baseline characteristic of each subject. The latter condition $Y^\tau(1)= Y^\tau(0)= 0$ ensures that causal treatment effects on $S$ are defined. With $h(x,y)$ being a known contrast function satisfying $h(x,y)=0$ if and only if $x =y$, GH defined the CEP surface as 
\begin{equation*}\label{eq1}
CEP^{risk}(s_1, s_0)\equiv h(risk_1(s_1, s_0),risk_0(s_1, s_0)).
\end{equation*}

The marginal CEP curve, closely related to the CEP surface, is also of great value in studying biomarkers as effect modifiers. It contrasts the risks averaged over the distribution of $S(0)$: $mCEP^{risk}(s_1)\equiv h(risk_1(s_1),risk_0(s_1))$, where $risk_z(s_1)\equiv P\left ( Y(z)=1|S(1)=s_1, Y^\tau(1)= Y^\tau(0)= 0\right )$. A example of the marginal CEP curve is vaccine efficacy as a function of $S(1)$, which is a causal estimand measuring the relative reduction in infection risk conferred by randomizing to vaccine versus placebo for different levels of $S(1)$: $VE(s_1)\equiv 1-\frac{risk_1(s_1)}{risk_0(s_1)}$.

In this manuscript, we propose methods to estimate the causal estimand for bivariate treatment effect modification analysis $mCEP^{risk}(S(1),B)\equiv h(risk_1(S(1),B),risk_0(S(1),B)$, applicable to both the BIP-only design and the BIP+CPV design based on an estimated likelihood approach in the presence of non-monotone missingness. Furthermore, in the special case where B denotes the baseline biomarker values, we also derive the estimator for the baseline seropositive mCEP curve ($mCEP^{risk}(S(1),B>c)\equiv h(risk_1(S(1),B>c),risk_0(S(1),B>c))
$) and the baseline seronegative mCEP curve ($mCEP^{risk}(S(1),B=c)\equiv h(risk_1(S(1),B=c),risk_0(S(1),B=c))$).

We make the common assumptions for randomized clinical trials of SUTVA (A1), ignorable treatment assignment (A2), and Equal early clinical risk: $P\left ( Y^\tau(1)=Y^\tau(0) \right )=1$ (A3). These three assumptions reduce the number of missing potential outcomes and help with identifiability of our estimands. They have been used and discussed in details in previously literature (\citet{gilbert2008evaluating},\citet{gabriel2013evaluating}, \citet{huang2011comparing},\citet{huang2013design}). Henceforth, we drop the notation of $Y^\tau(1)=Y^\tau(0) =0$ and tacitly assume all probabilities condition on $Y^\tau(1)=Y^\tau(0) =0$. Furthermore, we assume the risk functions have a generalized linear model form: $risk_{z}\left \{ S(1),B,X \right \}=g\left \{ \beta;S(1),B,Z,X \right \}$ for some known link function $g(\cdot )$, for $z=0,1$ (A4).

In order to replace the unobservable $S^*(1)$ among placebo recipients with the closeout measurement $S^*_c$, the following two assumptions are made for the BIP+CPV design only:
\begin{enumerate}
\item[(A5)] Time constancy of immune response: For event-free  placebo recipients, $S^*(1)=S^{* \rm{true}}+U_1$, and $S^*_c=S^{* \rm{true}}+U_2$, for some underlying $S^{* \rm{true}}$ and i.i.d. measurement errors $U_1$, $U_2$ that are independent of one another. 
\item[(A6)] No placebo subjects event-free at closeout experience the endpoint over the next $\tau$ time-units. 
\end{enumerate}

Henceforth we consolidate the notation and let $S^*$ be the potential outcome of $S^*$ under treatment arm $Z=1$, either obtained during the standard trial follow-up for vaccine recipients or replaced by the CPV measurements for placebo recipients. We let $S\equiv max(S^*,c)$ and $\delta$ to be the indicator that $S$ is measured. In addition, if B denotes the baseline biomarker values, we also replace a missing $B$ with $S(0)$ if it is available for placebo recipients based on (A3) and the next assumption (A7):
\begin{enumerate}
\item[(A7)] $B^*=B^{* \rm{true}}+U_3$, and $S^*(0)=B^{* \rm{true}}+U_4$, for some underlying $B^{* \rm{true}}$ and i.i.d. measurement errors $U_3$, $U_4$ that are independent of one another. 
\end{enumerate}

Henceforth, if $B$ denotes the baseline biomarker values subject to detection limit: $B= max(B^*,c)$, then we assume that $B^*$ denotes the baseline biomarker values that could potentially being replaced by $S^*(0)$. We let $B\equiv max(B^*,c)$ and $\delta_B$ to be the indicator that $B$ is available.

For the settings we consider in this article, $\left \{ i: \delta_i=1 \right \}$ and $\left \{ i: \delta_{Bi}=1 \right \}$ do not need to hold an inclusion relationship. In section \ref{sec4}, we discuss the special cases where $\delta=1$ implies $\delta_B=1$.

Lastly, we assume observed data $O_i\equiv (Z_i,X_i,\delta_{i},\delta_{i}S_i,\delta_{Bi},\delta_{Bi}B_i,Y^\tau_i,Y_i)', i=1,...,n$ are independent and identically distributed (i.i.d).

\subsection{Risk Model Parameters Estimation}
We propose an estimated likelihood estimator based on conditional likelihood for our risk model parameters $\beta$. Subjects with $\delta_{Bi}=\delta_{i}=1$ contribute to likelihood $risk_{Z_i}(S_i,B_i,X_i; \beta)^{Y_i}(1-risk_{Z_i}(S_i,B_i,X_i; \beta))^{1-Y_i}$. The likelihood contribution for subjects with $\delta_{Bi}=1$ and $\delta_{Si}=0$ is obtained by integrating $risk_{Z_i}(\cdot ,B_i,X_i; \beta)$ over the conditional cdf $F^{S|B,X}$. The contribution for subjects with $\delta_{i}=1$ and $\delta_{Bi}=0$ is obtained by integrating $risk_{Z_i}(S_i,\cdot,X_i; \beta)$ over the conditional cdf $F^{B|S,X}$. The contribution for subjects with $\delta_{i}=\delta_{Bi}=0$ is obtained by integrating $risk_{Z_i}(\cdot,\cdot,X_i; \beta)$ over the conditional cdf $F^{(B,S)|X}$. Define nuisance parameter $\nu  \equiv  \left ( F^{B|X}, F^{S|B,X}, F^{(B|S,X} \right )$. Then the condition likelihood is $$L(\beta,\nu)\equiv \prod_{i=1}^{n}f(Y_i|Z_i,X_i,  \delta_{Bi}, \delta_{i},\delta_{Bi}B_i, \delta_{i}S_i)$$ where

\begin{align*}
&f(Y|Z,X,  \delta_{B}, \delta,\delta_{B}B, \delta S)\\
=&\left \{ risk_{Z}(S,B,X; \beta)^{Y}(1-risk_{Z}(S,B,X; \beta))^{1-Y} \right \}^{\delta_B\delta}\\
\times & \left \{   \left ( \int risk_{Z}(s_1,B,X; \beta) dF^{S|B,X}(s_1|B,X)  \right ) ^Y \right. \\
&\qquad \times \left.  \left ( 1-\int risk_{Z}(s_1,B,X; \beta) dF^{S|B,X}(s_1|B,X)  \right ) ^{1-Y}  \right \} ^{\delta_B (1-\delta)}\\
\times & \left \{   \left ( \int risk_{Z}(S,b,X; \beta) dF^{B|S,X}(b|S,X)  \right ) ^Y \right. \\
&\qquad \times \left.  \left ( 1-\int risk_{Z}(S,b,X; \beta) dF^{B|S,X}(b|S,X)  \right ) ^{1-Y}  \right \} ^{(1-\delta_B)\delta}\\
\times &\left \{   \left ( \int \int risk_{Z}(s_1,b,X; \beta) dF^{S|B,X}(s_1|b,X) dF^{B|X}(b|X)  \right ) ^Y\right. \\
&\qquad \times \left. \left ( 1-\int \int risk_{Z}(s_1,b,X; \beta) dF^{S|B,X}(s_1|b,X) dF^{B|X}(b|X)  \right ) ^{1-Y}  \right \} ^{(1-\delta_B)(1-\delta)}
\end{align*}

We consider the estimated likelihood approach by Pepe and Fleming (1991) \citet{pepe1991nonparametric} where consistent estimates of $\nu$ are obtained first and then $L(\beta,\hat{\nu})$ is maximized in $\beta$. Here we assume $F^{B|X}$ and $F^{S|B,X}$ have particular parametric distribution. For example, we might assume $F^{B|X}$ is censored normal and $F^{S|B,X}$ is also censored normal, with left-censoring of values below c. Then according to Bayes' theorem, we have $f(B|S,X)=\frac{f(S|B,X)f(B|X)}{\int f(S|b,X)\cdot f(b|X)db}$. We obtain the maximum likelihood estimator (MLE) for $F^{B|X}$ using data from all individuals with B measured, $\left \{ i: \delta_{Bi}=1 \right \}$. For estimation of $F^{S|B,X}=F^{S|B,X,Z=1}$, we use data from vaccine recipients who have both $S$ and $B$ measured, with inverse probability weighting (IPW) used to account for biased sampling of $S$. Even if there is a CPV component in the study design, we can not use $S$ for placebo recipients obtained during CPV because placebo recipients who are infected at study closeout have zero probability of obtaining $S$ thus IPW is not applicable. The estimator of $\beta$ is then derived as the maximizer of the estimated likelihood $L(\beta,\hat{\nu})$ and $\widehat{mCEP}(S,B,X)=h(g\left \{ \widehat{\beta};S,B,Z=1,X \right \},g\left \{ \widehat{\beta};S,B,Z=0,X \right \}$ provides an estimate of bivariate treatment effect modification by $S$ and $B$, adjusting for $X$. Standard errors for $\widehat{\beta}$ can be estimated using a perturbation resampling technique. In essence, one can generate $n$ random realizations of $\epsilon $ from a known distribution with mean of 1 and variance of 1 to create $\mathcal{E}\equiv \left \{   \epsilon_i, i=1,2,...,n\right \}$. Let $L^{(\epsilon)}(\beta,\hat{\nu}^{(\epsilon)})$ be a perturbed version of $L(\beta,\hat{\nu})$, where $L^{(\epsilon)}(\beta,\nu)\equiv \prod_{i=1}^{n}f(Y_i|Z_i,X_i,  \delta_{Bi}, \delta_{i},\delta_{Bi}B_i, \delta_{i}S_i)\cdot \epsilon_i$ and $\nu^{(\epsilon)}$ is the perturbed estimator of $\nu$ with $\mathcal{E}$ being the weights. Then the perturbed estimator $\beta^{(\epsilon)}$ is a maximized of $L^{(\epsilon)}(\beta,\hat{\nu}^{(\epsilon)})$. In practice, one may obtain a variance estimator of $\hat{\beta}$ based on the empirical variance of B realizations of $\beta^{(\epsilon)}$. In our simulation and example, we use B=500.

\subsection{Baseline Seropositive/Seronegative mCEP Curves} 
In this section, we study the special case where B denotes the baseline biomarker values subject to detection limit, c, and derive the estimator for the marginal mCEP curve ($mCEP(S)\equiv h(risk_1(S),risk_0(S))$), baseline seropositive mCEP curve ($mCEP(S,B>c)\equiv h(risk_1(S,B>c),risk_0(S,B>c))$) and the baseline seronegative mCEP curve ($mCEP(S,B=c)\equiv h(risk_1(S,B=c),risk_0(S,B=c))$). With some calculations, the risk functions in our estimands of interest can be expressed as: marginal risk function $risk_Z(S,X)=\int_{c-}^{\infty }risk_Z(S,b,X)dF^{B|S,X}(b|S,X)$; seropositive risk function $risk_Z(S,B>c,X)=\frac{P(Y(Z)=1,B>c|S,X)}{P(B>c|S,X)}=\frac{ \int_{c+}^{\infty }  risk_Z(S,b,X) dF^{B|S,X}(b|S,X) }{P(B>c|S,X)}$; and seronegative risk function $risk_Z(S,B=c,X)$. All three risk functions can be estimated based on $\hat{\beta}$ and $\hat{\nu}$.

We consider situations where $X$ is categorical with $D$ levels: $x_1, x_2,...,x_D$. Then 
\begin{align}
risk_Z(S) &=\sum_{j=1}^{D} risk_Z(S,x_j) \cdot P(X=x_j|S) \label{BLsero:MargianlRisk}\\
risk_Z(S,B>c) & =\sum_{j=1}^{D} risk_Z(S,B>c,x_j) \cdot P(X=x_j|S,B>c) \label{BLsero:BLposRisk}\\
risk_Z(S,B=c) &=\sum_{j=1}^{D} risk_Z(S,B=c,x_j) \cdot P(X=x_j|S,B=c). \label{BLsero:BLnegRisk}
\end{align}

We model $P(X|S)$, $P(X|S, B>c)$, and $P(X|S, B=c)$ using a multinomial logistic function with parameter $\gamma$ and estimate $\gamma$ by MLE. Because sampling of $B$ and $S$ depends on other phase-I variables such as $Y$, inverse prob. weighting (IPW) (Horvitz and Thompson, 1952) can be implemented.

Appendix provides a detailed estimation procedure of $mCEP(S)$, $mCEP(S,B>c)$, and $mCEP(S,B=c)$ for the case where $F^{B|X}$ is assumed censored normal, $F^{S|B,X}$ is assume censored normal, and the risk functions take the form $risk_{z}\left \{ S,B,X \right \}=g\left \{ \beta;S,B,Z,X \right \}=\Phi(\beta_0+\beta_1Z+\beta_2S+\beta_3Z\cdot S+\beta_4B+\beta_5Z\cdot B+\beta_6X)$. 

A perturbation resampling method can be used to make simultaneous inference of marginal mCEP curve, baseline seropositive mCEP curve and baseline seronegative mCEP curve. To be specific, perturbed estimators $\beta^{(\epsilon)}$ and $\nu^{(\epsilon)}$ are obtained based on $\mathcal{E}$. Then the corresponding perturbed estimators $\widehat{mCEP}^{(\epsilon)}(S)$, $\widehat{mCEP}^{(\epsilon)}(S,B>c)$, and $\widehat{mCEP}^{(\epsilon)}(S,B=c)$ are obtained by plugging in $\beta^{(\epsilon)}$ and $\nu^{(\epsilon)}$ in equation \ref{BLsero:MargianlRisk}, \ref{BLsero:BLposRisk}, and \ref{BLsero:BLnegRisk}. Repeat this process $B$ times to obtain $B$ realizations of $\widehat{mCEP}^{(\epsilon)}(S)$, $\widehat{mCEP}^{(\epsilon)}(S,B>c)$, and $\widehat{mCEP}^{(\epsilon)}(S,B=c)$, and calculate the sample standard deviations $\hat{\sigma}_{mCEP}(S)$, $\hat{\sigma}_{mCEP}(S, B>c)$, $\hat{\sigma}_{mCEP}(S, B=C)$. $100(1-\alpha)\%$ pointwise confidence intervals can be constructed as 
\begin{align*}
\widehat{mCEP}(S) & \pm \mathcal{Z}_{1-\alpha/2} \hat{\sigma}_{mCEP}(S)\\
\widehat{mCEP}(S, B>c) & \pm \mathcal{Z}_{1-\alpha/2} \hat{\sigma}_{mCEP}(S, B>c)\\
\widehat{mCEP}(S, B=c) & \pm \mathcal{Z}_{1-\alpha/2} \hat{\sigma}_{mCEP}(S, B=c).
\end{align*}
And $100(1-\alpha)\%$ simultaneous confidence bands for $ S \in  \zeta$ can be constructed as 
\begin{align*}
\widehat{mCEP}(S) & \pm \mathcal{Q}_{1-\alpha} \hat{\sigma}_{mCEP}(S)\\
\widehat{mCEP}(S, B>c) & \pm \mathcal{Q}'_{1-\alpha} \hat{\sigma}_{mCEP}(S, B>c)\\
\widehat{mCEP}(S, B=c) & \pm \mathcal{Q}''_{1-\alpha} \hat{\sigma}_{mCEP}(S, B=c),
\end{align*}
where $\mathcal{Z}_{1-\alpha/2}$ is the $100(1-\alpha/2)$th percentile of $N(0,1)$, $\mathcal{Q}_{1-\alpha}$ is the $100(1-\alpha)$th percentile of $sup_{S \in \zeta}  \left |\frac{ \sqrt{n}\left \{ \widehat{mCEP}^{risk(\epsilon)}(S)- \widehat{mCEP}(S)\right \} }{\hat{\sigma}_{mCEP}(S)}  \right | $, and $\mathcal{Q}'_{1-\alpha}$ and $\mathcal{Q}''_{1-\alpha}$ defined similar to $\mathcal{Q}_{1-\alpha}$ with the mCEP estimator, mCEP perturbed estimator and standard error estimator replaced by its own version. 

Furthermore, simultaneous inference enables evaluation of the hypothesis testing of $H_0$: $mCEP(S,B>c)=mCEP(S,B=c)$ for $S \in \zeta$. We first construct the simultaneous confidence band for $\widehat{mCEP}(S, B>c) -\widehat{mCEP}(S, B=c) $. Let $\widehat{\sigma}(\widehat{mCEP}(S, B>c) -\widehat{mCEP}(S, B=c))$ denote the sample standard deviation of the perturbed estimates $\widehat{mCEP}^{(\epsilon)}(S, B>c) -\widehat{mCEP}^{(\epsilon)}(S, B=c)$. Let $\mathcal{Q}'''_{1-\alpha}$ be the $100(1-\alpha)$th percentile of $$sup_{S \in \zeta}  \left |\frac{ \sqrt{n}\left \{ \widehat{mCEP}^{(\epsilon)}(S,B>c)-\widehat{mCEP}^{(\epsilon)}(S,B=c)- \left (\widehat{mCEP}(S,B>c)-\widehat{mCEP}(S,B=c)  \right )\right \} }{\widehat{\sigma}\left (\widehat{mCEP}(S, B>c) -\widehat{mCEP}(S, B=c)  \right )}  \right | .$$Subsequently, the $100(1-\alpha)\%$ simultaneous confidence bands for $\widehat{mCEP}(S, B>c) -\widehat{mCEP}(S, B=c) $, $ S \in  \zeta$ is $$\left ( l_\alpha(S),u_\alpha(S) \right )\equiv \widehat{mCEP}(S, B>c) -\widehat{mCEP}(S, B=c) \pm \mathcal{Q}'''_{1-\alpha} \widehat{\sigma}(\widehat{mCEP}(S, B>c) -\widehat{mCEP}(S, B=c)).$$ The two-sided p-value for the testing $H_0$: $mCEP(S,B>c)=mCEP(S,B=c)$ is defined as the minimum of $\alpha_1$ and $\alpha_2$ that satisfy
$$\begin{matrix}
\rm{inf}_{S \in \zeta} u_{\alpha_1}(S)=0,  & \rm{sup}_{S \in \zeta} l_{\alpha_1}(S)\leq 0\\ 
 \rm{sup}_{S \in \zeta} l_{\alpha_2}(S)=0, & \rm{inf}_{S \in \zeta} u_{\alpha_2}(S)\geq  0
\end{matrix}.$$
Note that at least one of $\alpha_1$ and $\alpha_2$ always exists.

\section{Simulation Studies}


Through simulation studies, we evaluate the finite-sample performance of our proposed estimators. Simulation data are generated with 10,000 subjects randomized to vaccine and placebo by a ratio of 2:1. Baseline covariate $X$ was generated with a multinomial distribution to have four categories, 1, 2, 3, and 4 with corresponding probabilities of 0.25, 0.25, 0.25 and 0.25. $X_2$, $X_3$, and $X_4$ are dummy variables indicating category 2, 3, or 4, respectively. Baseline biomarker values $B$ were generated from a normal distribution with mean of $1.38+0.93X_2+1.25X_3-0.25X_4$ and standard deviation of 0.86. $S$ were generated from a normal distribution with mean of $1.5+0.5B+0.2X_2-0.1X_3+0.4X_4$ and standard deviation of 0.4, which indicates a correlation of 0.7 between $S$ and $B$. Let the limit of biomarker value detection be 1. Simulated values of $S$ and $B$ less than 1 were set equal to 1. We assume a probit risk model of the clinical outcome $Y$ conditional on $S$, $B$, $Z$, and $X$: $P(Y=1|S,B,Z,X)=\Phi(\beta_0+\beta_1Z+\beta_2S+\beta_3Z\cdot S+\beta_4B+\beta_5Z\cdot B+\beta_6X)$. We set $(\beta_0, \beta_1, \beta_2, \beta_3, \beta_4, \beta_5, \beta_6)$ as $\left (-0.50,  0.16, -0.34, -0.21, -0.25, 0,   \left (0.24,  0.11,  0.20  \right )  \right )$ so that the probability of infection equals 0.04 in the placebo arm and 0.02 in the vaccine arm. These simulation parameters were chosen to reflect the characteristics of the two Phase 3 Dengue trials. To achieve a non-monotone sampling design, 35\% of study participants have $B$ retained. For the BIP-only design, $S$ is set missing for all placebo recipients and retained in all cases and all subjects with $B$ measured in the vaccine arm, that is $\left \{ i: Z_i=1, Y_i=1 \right \}\cup \left \{ i: Z_i=1, \delta_{Bi}=1 \right \}$. For the BIP+CPV design, 70\% of event-free placebo recipients are included in the CPV component and have $S$ retained. Simulation results are based on 500 Monte-Carlo simulations and for each simulation 250 perturbation iterations are generated to construct point-wise confidence intervals and simultaneous confidence bands.

\begin{figure}[!p]
\centering\includegraphics[scale=0.7]{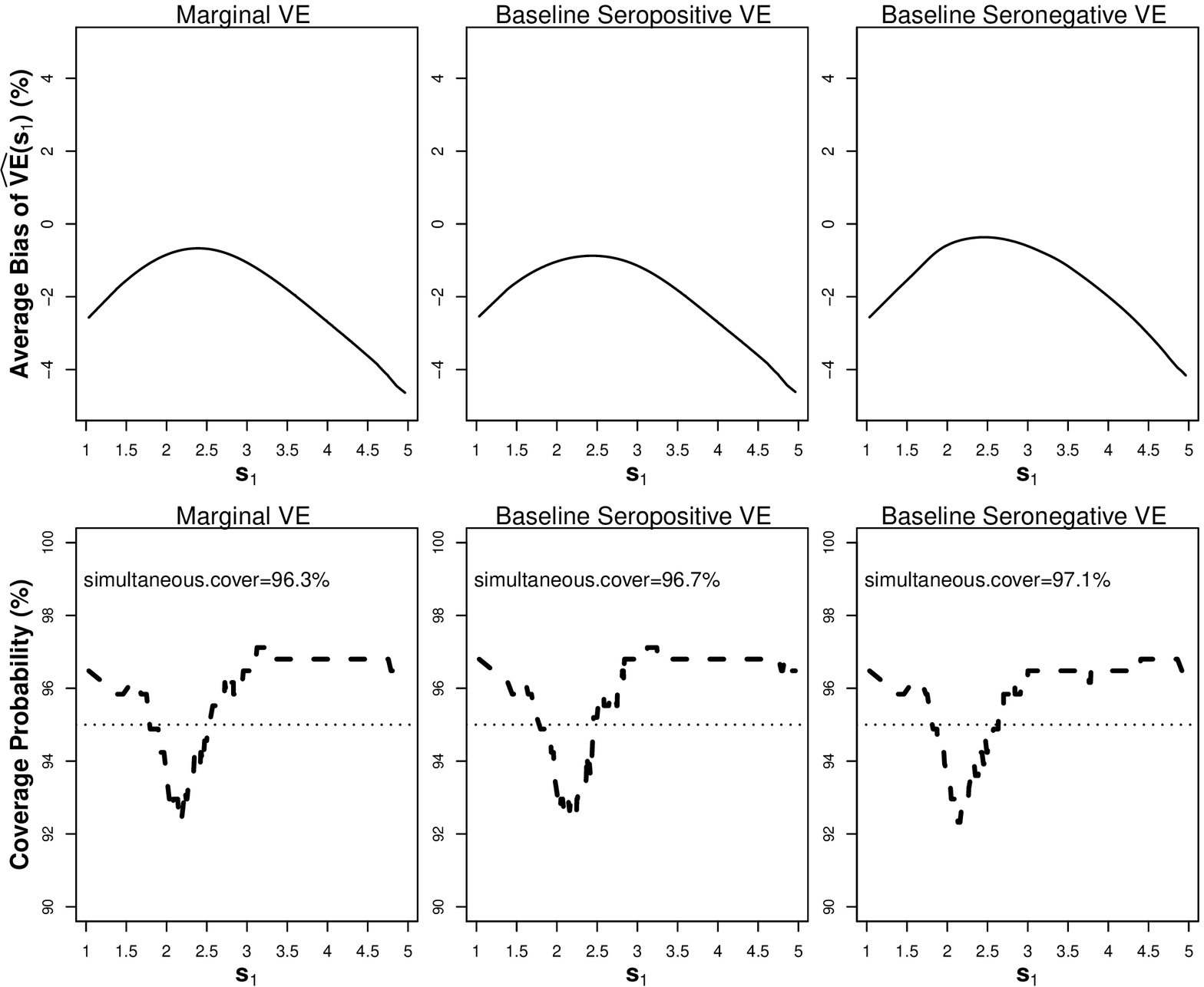}
\caption{Average bias for our proposed estimators ${\rm \widehat{VE}}\left (S(1)=s_1\right )$, ${\rm \widehat{VE}}\left (S(1)=s_1, B>c  \right )$ and ${\rm \widehat{VE}}\left (S(1)=s_1, B=c  \right )$ and coverage probabilities of 95\% perturbation Wald confidence intervals in a BIP-only design.}
\label{Fig:BLseroVEcover_BIP}
\end{figure}
\begin{figure}[!p]
\centering\includegraphics[scale=0.7]{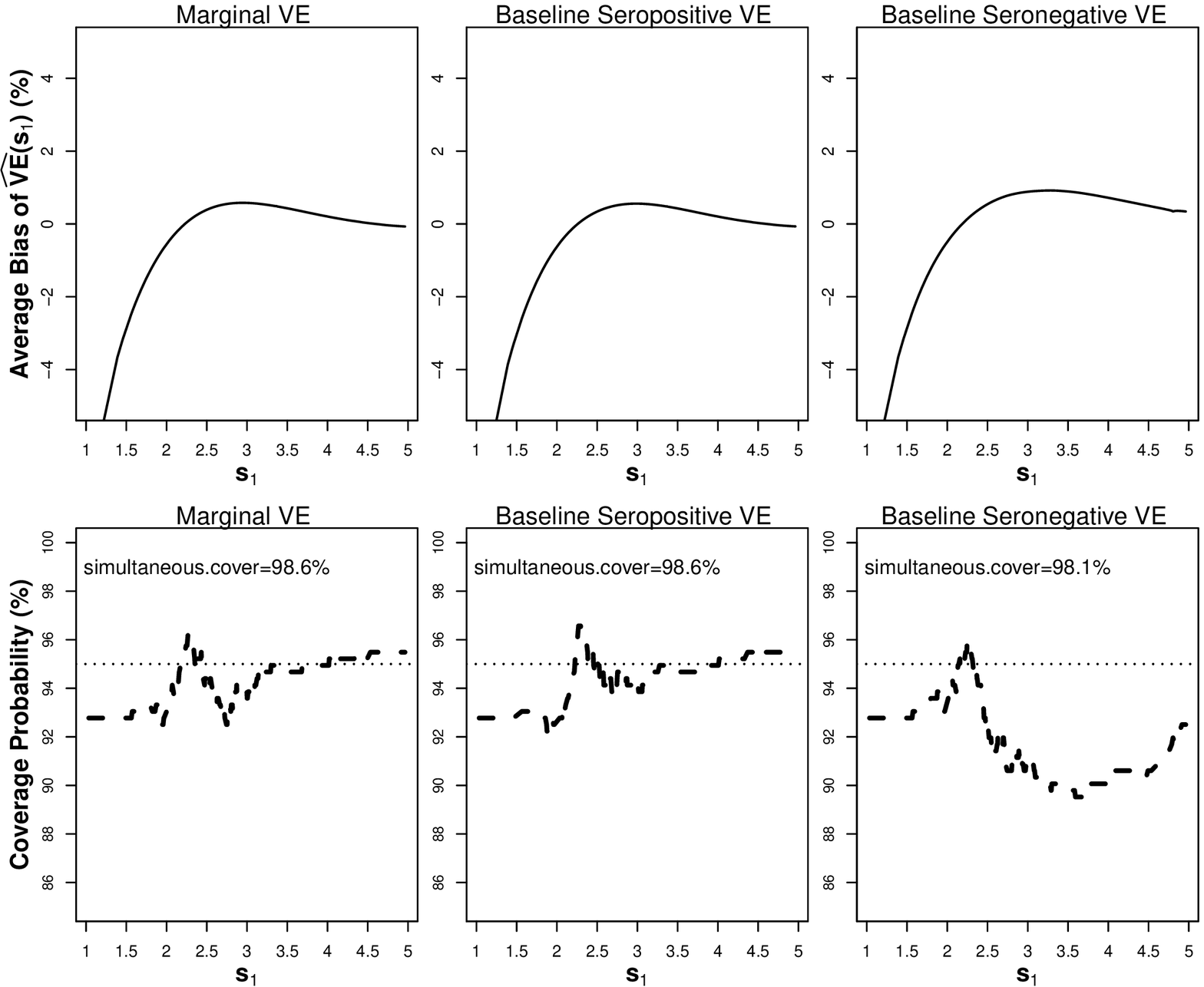}
\caption{Average bias for our proposed estimators ${\rm \widehat{VE}}\left (S(1)=s_1\right )$, ${\rm \widehat{VE}}\left (S(1)=s_1, B>c  \right )$ and ${\rm \widehat{VE}}\left (S(1)=s_1, B=c  \right )$ and coverage probabilities of 95\% perturbation Wald confidence intervals in a BIP+CPV design.}
\label{Fig:BLseroVEcover_CPV}
\end{figure}

We then evaluate the finite-sample performance of our proposed estimators for the marginal VE curve (${\rm VE}\left (S=s_1\right )$), baseline seropositive VE curve (${\rm VE}\left (S=s_1, B>c  \right )$), and baseline seronegative VE curve (${\rm VE}\left (S=s_1, B=c  \right )$). Results are presented in Figure~\ref{Fig:BLseroVEcover_BIP} for BIP-only design and Figure~\ref{Fig:BLseroVEcover_CPV} for BIP+CPV design. The empirical coverage levels of the 95\% simultaneous confidence bands from the perturbation methods are also reported as "simultaneous.cover" in Figure~\ref{Fig:BLseroVEcover_BIP} and \ref{Fig:BLseroVEcover_CPV}. They demonstrate satisfactory performance of our proposed estimators, including nominal coverage probabilities of the confidence intervals given fixed $s_1$ values and simultaneous confidence band across all $s_1$ values. 

\section{Application to the CYD14 and 15 Trials}
CYD14 (CYD15) is an observer-masked, randomized controlled, multi-center, phase 3 trial in five countries in the Asia-Pacific (Latin America) region where participants were randomized to receive three injections at month 0, 6, and 12. The primary goal is to assess vaccine efficacy against symptomatic, virologically confirmed dengue (VCD) occurring more than 28 days after the third injection. CYD14 achieved 56.5\% efficacy (95\% CI 43.8-66.4) and CYD15 achieved 60.8\% efficacy (95\% CI 52.0-68.0) in the per-protocol population. Concentrations of dengue neutralizing antibody titers to each of the four dengue serotype strains at month 13 were measured for all VCD cases and a subset of controls, of which only a fraction have their baseline titers measured. In this illustration, we applied our proposed method to data pooling across CYD14 and CYD15 9-16 year olds to assess how VE varied by Month 13 titers within baseline seropositive and seronegative subgroups. \citet{moodie2017denguecorcop} provided the justification for pooling data across these two trial for data analysis purpose. We let $S$ be the average of the log10-transformed neutralizing antibody titers to each of the four dengue serotypes at month 13 and $B$ be the average of the log10-transformed titers at baseline. Baseline seropositive and seronegative subgroups are defined as $B>1$ and $B=1$. CYD14 and CYD15 hold a non-monotone sampling design. $B$ is available for everyone in the immunogenicity subset ($\left \{ i: \delta_B=1 \right \}$) and $S$ is available for all vaccine recipients who are either in the immunogenicity subset or are cases ($\left \{ i:Z_i=1,  \delta_{Bi}=1 \right \}\cup \left \{ i: Z_i=1, Y_i=1 \right \}$). These two sets do not have an inclusion relationship. $X$ denotes participants' age and country categories, and Y is the indicator of VCD that took place between month 13 and end of follow up (month 25).

\begin{figure}[!p]
\centering\includegraphics[scale=0.7=8]{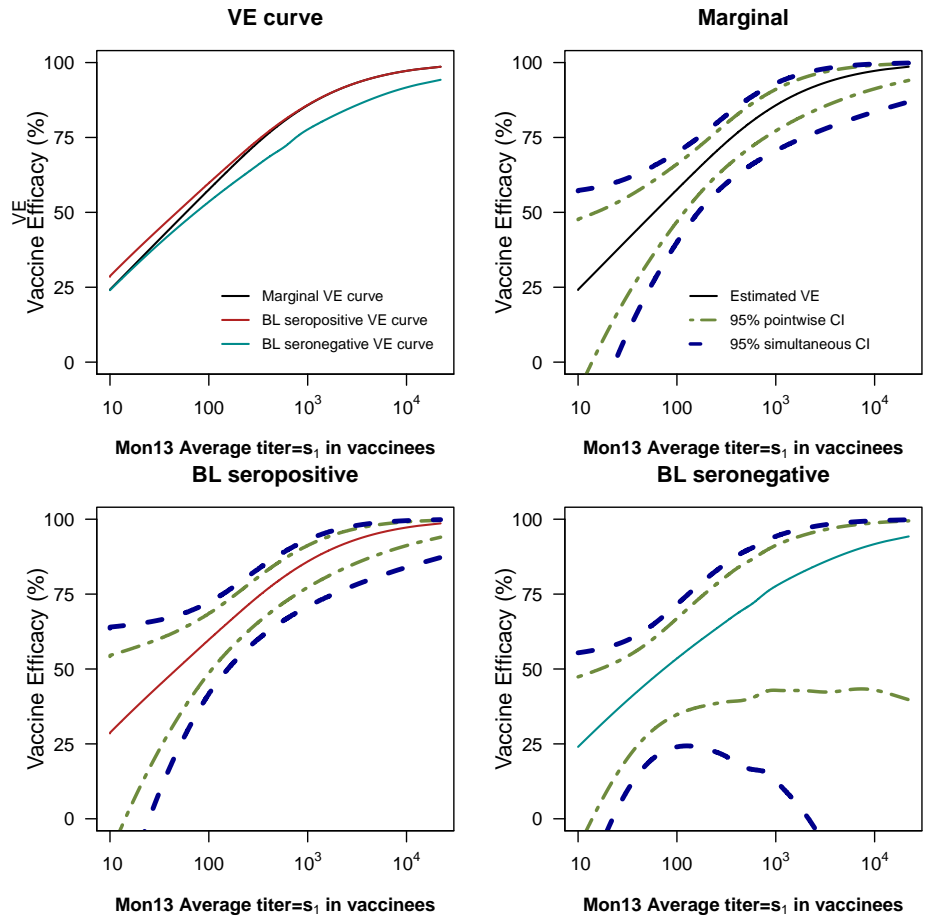}
\caption{Estimated vaccine efficacy by average $\rm{log}_{10}$ titer at Month 13 with 95\% pointwise confidence intervals and simultaneous confidence bands in CYD14
and CYD15 9-16-year-olds.}
\label{Fig:BLseroDengueVE}
\end{figure}

Figure \ref{Fig:BLseroDengueVE} shows the estimated VE curve and 95\% CIs and CBs based on 500 perturbation iterations. VE curves were similar for baseline seropositive and baseline seronegative subgroups, with estimated VE approximately 25\% for vaccine recipients with no seroresponse at Month 13. For vaccine recipients with Month 13 average titers of 500 and 10,000, estimated VE was 79.3\% and 97.3\% for the baseline seropositive subgroup compared to 70.4\% and 91.8\% for the baseline seronegative subgroup, respectively. Furthermore, we tested the null hypothesis $H_0$: BL seropositive VE$(S)=$BL seronegative VE$(S)$ for $S \in $ range of month 13 average titer in vaccinees in the data using procedure provided in section 2.2, which gave a p-value of 0.35. This suggests that the seropositive VE curve was not significantly different from the seronegative VE curve, implying that it is not the new neutralization response generated by the vaccine over baseline value that predicts VE, but rather the absolute titer achieved following vaccination. See \citet{moodie2017denguecorcop} for the reporting of the full analysis to a clinical audience.

\section{Discussion}
\label{sec4}
In this article, we developed an estimated likelihood approach to evaluate the bivariate treatment effect modification analysis by biomarker-based principal strata and baseline covariates in general settings without requirements of a nested sub-sampling relationship between the immune response biomarker and baseline predictors suitable to both the BIP-only design and the BIP+CPV design. In our settings, the biomarker sampled set $\left \{ i: \delta_{i}=1 \right \}$ and the baseline covariates sampled set $\left \{ i: \delta_{Bi}=1 \right \}$ do not need to be a subset of the other. Dengue vaccine trials CYD14 and CYD15 are examples of this non-inclusion relationship. 

Our proposed method can apply to the special case where $\left \{ i: \delta_{i}=1 \right \}$ is a subset of $\left \{ i: \delta_{Bi}=1 \right \}$. An example would be a three-phase sampling design when lab-assay-based baseline covariates are only measured from a subset of the trial participants due to high costs of acquiring lab assay and the vaccine-induced immune response is further measured among a subset of participants for whom the lab-assay-based baseline covariates are available. The phase 3 Zostavax Efficacy and Safety Trial (ZEST) adopted such a three-phase sampling design to study the effect of the Zostavax vaccine against varicella zoster virus (VZV) (\citet{schmader2012efficacy}). Under this sampling framework, \citet{huang2017evaluating} proposed a semiparametric pseudo-score estimator based on conditional likelihood and also develop several alternative semiparametric estimated likelihood estimators when $B$ is discrete. One can think of our work as an extension of \citet{huang2017evaluating} that our proposed method can incorporate broader sampling settings including the one in Huang (2017) and our baseline covariate $B$ can be either discrete or continuous, possibly subject to detection limit left censoring. In general, our methods are applicable to intervention studies where a bivariate effect modification analysis is of interest where the bivariate is a post-randomization measurement and a baseline covariate, and measurements of one do not necessarily imply measurements of the other.

\section{Supplementary Material}
\label{sec6}

Supplementary material is available online at
\url{http://biostatistics.oxfordjournals.org}.

\section*{Acknowledgments}
The authors thank the participants, investigators, and sponsors of the CYD14 and CYD15 trials. Research reported in this publication was supported by Sanofi Pasteur and the National Institute of Allergy and Infectious Diseases (NIAID), National Institutes of Health (NIH), Department of Health and Human Services, under award number R37AI054165. The content is solely the responsibility of the authors and does not necessarily represent the official views of the NIH or Sanofi Pasteur.
{\it Conflict of Interest}: None declared.

\bibliographystyle{biorefs}
\bibliography{refs}

\end{document}